\documentclass[allclo]{FBSart}

\usepackage{amsfonts}
\usepackage{amssymb}
\usepackage[dvips,final]{graphicx}

\usepackage{cite}                         
\usepackage{multirow}                     
\usepackage{amsmath}
\usepackage{xspace}                       
\newcommand{\PLB}{\textnormal{Phys.\ Lett.\ }\textbf{B}}

\newcommand{\MeV}{\ensuremath{\mathrm{MeV}}}

\newcommand{\EFTNoPion}{EFT(${\pi\hskip-0.55em /}$)\xspace}

\newcommand{\NXLO}[1]{N\ensuremath{{}^{#1}}LO\xspace}

\newcommand{\wave}[3]{\ensuremath{{}^{#1}\mathrm{#2}_{#3}}}

\renewcommand{\Re}{\mathrm{Re}}
\renewcommand{\Im}{\mathrm{Im}}

\newcommand{\de}{\partial}

 \newcommand{\calK}{\mathcal{K}}

\title{Estimating Consequences of 3-Body Forces}

\author{Harald W.~Grie\3hammer\thanks{\mbox{\textit{E-mail address:} hgrie@gwu.edu;
    29 Nov 2007; arXiv:0711.4522. To appear in Few-Body Systems.}}  }

\institute{Center for Nuclear Studies, 
  The George Washington University, Washington DC 20052, USA}

\runningauthor{Harald W.~Grie\3hammer}

\runningtitle{Estimating Consequences of 3-Body Forces}

\begin{document}
\maketitle
\begin{abstract}
  Classifying the strengthes of three-body forces 3BFs with the condition that
  observables must be cut-off independent, i.e.~renormalised at each order,
  leads to surprising results with relevance for example for thermal neutron
  capture on the deuteron.  Details and a better bibliography in
  Ref.~\cite{Griesshammer:2005ga}.
\end{abstract}


Adding 3-Body Forces (3BFs) \emph{a posteriori} when theory and data disagree
is untenable when predictions are required. Effective Field Theories (EFTs),
see e.g.~\cite{Braaten:2004rn} for reviews, provide a model-independent way to
estimate their typical strength. For systems of three identical particles in
which short-range forces produce shallow two-particle bound states, and in
particular for the ``pion-less'' EFT of Nuclear Physics \EFTNoPion,
consistency arguments from renormalisation lead to a power-counting, namely a
recipe to systematically estimate the typical size of 3BFs in all partial
waves and orders, including external currents. 


\begin{figure}[!htb]
\begin{center}
  \includegraphics*[width=0.8\textwidth]{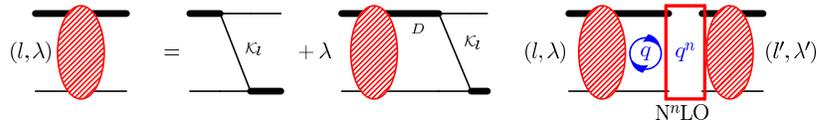}
  \caption{Left: integral equation of nucleon-deuteron scattering. Right:
    generic loop correction (rectangle) at \NXLO{n}. Thick line ($D$): $NN$
    propagator; thin line ($\calK_l$): propagator of the exchanged nucleon;
    ellipse: LO half off-shell amplitude.}
\label{fig:kinematics}
\label{fig:higherorders}
\end{center}
\end{figure}
We start from the Faddeev equation, Fig.~\ref{fig:kinematics}, in the $l$th
partial wave of the spin doublet ($\lambda=1$) and quartet
($\lambda=-\frac{1}{2}$) channels of nucleon-deuteron scattering at leading
order LO. The $NN$ amplitude is given by the leading term of the
Effective-Range Expansion. The 3-nucleon amplitude converges for large half
off-shell momenta $p$ as $p^{-s_0-1}$, with $s_0$ the solution to the
algebraic equation
\begin{equation}
  \label{eq:s}
  1=\left(-1\right)^l\;\frac{2^{1-l}\lambda}{\sqrt{3\pi}}\;
  \frac{\Gamma\left[\frac{l+s+1}{2}\right]\Gamma\left[\frac{l-s+1}{2}\right]}
  {\Gamma\left[\frac{2l+3}{2}\right]}\;
  {}_2F_1\left[\frac{l+s+1}{2},\frac{l-s+1}{2};
    \frac{2l+3}{2};\frac{1}{4}\right].
\end{equation}   
The asymptotics depends thus crucially on $l$ and the spin-isospin factor
$\lambda$ only.

As a description in terms of point-like interactions becomes inadequate beyond
a maximum momentum, low-energy observables must be insensitive to details of
the amplitude at large $p$, i.e.~to form and value of the regulator chosen.
This is the fundamental tenet of EFT: Include a 3BF \emph{if and only if}
needed to cancel cut-off dependence in low-energy observables, i.e.~as
counter-term for divergences which can not be absorbed by renormalising 2-body
interactions. 3BFs are thus not added out of phenomenological needs but to
guarantee that observables are insensitive to off-shell effects.  Counting
loop momenta in Fig.~\ref{fig:kinematics}, the \emph{superficial degree of
  divergence} of a higher-order correction is non-negative for
\begin{equation}
  \label{eq:twobodydivs}
  \Re[n-s_l(\lambda)-s_{l^\prime}(\lambda^\prime)]\geq0\;\;.
\end{equation}
The order $n$ at which a 3BF is needed is thus determined just by the order at
which a correction to the 3-body amplitude with only 2-body interactions
starts to depend on unphysical short-distance behaviour.

Let us re-visit the problem in position space. The Schr\"odinger equation for
the wave-function in the hyper-radial deuteron-nucleon distance $r$ at short
distances,
\begin{equation}
  \label{eq:hyperradial}
  \left[-\frac{1}{r}\;\frac{\de}{\de r}\;r\;\frac{\de}{\de r}+
  \frac{s_l^2(\lambda)}{r^2}-ME\right] F(r)=0\;\;,
\end{equation}
looks like the one for a free particle with centrifugal barrier. One would
thus expect $s_l\stackrel{?}{=}l+1$ (hyper-spherical co-ordinates!). It is
however well-known that the centrifugal term is attractive for three bosons
($\lambda=1$), contrary to expectations. The wave-function collapses to the
origin and seems infinitely sensitive to very-short-distance physics. In order
to stabilise the system -- or, equivalently, remove dependence on details of
the cut-off --, a 3BF must be added at LO.  This leads to the famous
``limit-cycle'' with its Efimov and Thomas effects and universal correlations,
see e.g.~Ref.~\cite{Braaten:2004rn} for reviews.

On the other hand, 3BFs are demoted if $s_l> l+1$ because the centrifugal
barrier provides \emph{more} repulsion than expected. The wave-function is
pushed further out and thus \emph{less} sensitive to short-distance
details.
%
About half of the 3BFs for $l\leq 2$ are \emph{weaker}, half \emph{stronger}
than one would expect simplistically, see Table~\ref{tab:ordering}. The higher
partial-waves follow expectation, as the Faddeev equation is then saturated by
the Born approximation.
\begin{table}[!ht]
\centerline{{\scriptsize
\begin{tabular}{|cc||c|c||c||c@{\hspace*{3pt}}c|}
  \hline
\multicolumn{2}{|c||}{partial-wave: \emph{in--out}}
&  na\"ive dim.~analysis &simplistic&&
\multicolumn{2}{|c|}{typ.~size}\\
  bosons& fermions&{$\mathrm{Re}[s_l(\lambda)+s_{l^\prime}(\lambda^\prime)]$} &
  {$l+l^\prime+2$}&&\multicolumn{2}{|c|}{if $Q^n\sim\frac{1}{3^n}$}\\[0.5ex]
  \hline\hline
  \wave{}{S}{}-\wave{}{S}{}&\wave{2}{S}{}-\wave{2}{S}{}&{LO}&
\NXLO{2}&{promoted}&{$100\%$}& ($10\%$)\\
  &\wave{2}{S}{}-\wave{4}{D}{}&\NXLO{3.1}&\NXLO{4}&promoted&{$3\%$}& ($1\%$)\\
\hline\hline
\wave{}{P}{}-\wave{}{P}{}&\wave{2}{P}{}-\wave{2}{P}{}&\NXLO{5.7}&\multirow{2}{2cm}{\hspace*{\fill}
\NXLO{4}\hspace*{\fill}}&demoted&0.2\%&\multirow{2}{0.8cm}{\hspace*{\fill}
($1\%$)\hspace*{\fill}}\\
&\hspace*{-1ex}\wave{2}{P}{}-\wave{2}{P}{},
\wave{4}{P}{}-\wave{4}{P}{}&\NXLO{3.5}&&promoted&{$2\%$} &\\
\hline\hline
&\wave{4}{S}{}-\wave{4}{S}{}&{\NXLO{4.3+2}}&{\NXLO{2+2}}&{demoted}&{$0.1\%$}&\multirow{2}{0.8cm}{\hspace*{\fill}
($1\%$)\hspace*{\fill}}\\
&\wave{4}{S}{}-\wave{2}{D}{}&\NXLO{5.0}&
\NXLO{4}
&demoted&{$0.4\%$}&
  \\ 
\hline
\multicolumn{2}{|c||}{higher}&{$\sim$ as simplistic}&{\NXLO{l+l^\prime+2}}&&&\\
\hline
  \end{tabular}}
}
\caption{Order of some leading 3BFs in nucleon-deuteron scattering, indicating if
  actual values from eqs.~(\ref{eq:s}/\ref{eq:twobodydivs}) are stronger
  (``promoted'')  or weaker (``demoted'') than the simplistic estimate. Last column:
  typical size of 3BF in \EFTNoPion; in parentheses size from the simplistic
  estimate.} 
  \label{tab:ordering}
\end{table}

Demotion might seem an academic dis-advantage -- to include some unnecessary
higher-order corrections does not improve the accuracy of the result.  But
demotion is pivotal when predicting the experimental precision necessary to
dis-entangle 3BFs in observables, and here the error-estimate of EFTs is
crucial. 

An example is the cross-section of triton radiative capture $nd \to t\gamma$
at thermal energies, see Ref.~\cite{withSadeghi} for details and references.
Nuclear models give a spread of $[0.49\dots0.66]\;\mathrm{mb}$, depending on
the 2-nucleon potential and inclusion of the
$\Delta(1232)$~\cite{competition}.  Restoring gauge-invariance reduces the
spread to $[0.52\dots0.56]\;\mathrm{mb}$, but the discrepancy to experiment
increases when gauge-invariant three-nucleon currents are added~\cite{competition}.  On the other
hand, this low-energy process should be insensitive to details of Physics at
$300\;\MeV$. Indeed, the power-counting of 3BFs applies equally with external
currents. No new 3BFs are needed up to \NXLO{2} to render cut-off
independence.  The result converges order by order,
\begin{equation}
   \sigma_\mathrm{tot}=[
   0.485(\mathrm{LO})+0.011(\mathrm{NLO})+0.007(\text{\NXLO{2}})
   ]\;\mathrm{mb}=
   [0.503\pm0.003]\;\mathrm{mb}\;\;,
\end{equation} 
is cut-off independent and compares well with experiment,
$[0.509\pm0.015]\;\mathrm{mb}$. In contradistinction to earlier potential
models, it is manifestly gauge-invariant.

The potential models reproduce the input of \EFTNoPion: the nucleon magnetic
moments, deuteron and triton binding energies, $NN$ and $nd$ scattering
lengthes, and the thermal cross-section of $np\to d\gamma$. That their results
vary dramatically is at odds with universality, a key aspect of EFTs: Answers
from models with the same input should agree with \EFTNoPion within the
projected accuracy. This conflict poses a puzzle whose resolution is not (yet)
clear.


With these findings, \EFTNoPion is a self-consistent field theory which
contains the minimal number of interactions at each order to be
renormalisable. Each 3-body counter-term gives rise to one
subtraction-constant, fixed by a 3-body datum.  This method is applicable to
any EFT with an infinite number of diagrams at LO, e.g.~because of shallow
bound-states.  It leads at each order and to the prescribed level of accuracy
to a cut-off independent theory with the smallest number of experimental
input-parameters.  The power-counting is thus not constructed by educated
guesswork but by investigations of the renormalisation-group properties of
couplings and observables using rigorous methodology.

\begin{acknowledge}
  It is a pleasure to thank the organisers for creating a highly stimulating
  conference and a warm welcome. Supported in part by the US Department of
  Energy (DE-FG02-95ER-40907) and a CAREER-grant PHY-0645498 of the US
  National Science Foundation.
\end{acknowledge}

\end{document}